\documentclass[twocolumn,superscriptaddress,showpacs,preprintnumbers,amsmath,amssymb,prl]{revtex4}
\usepackage[dvips]{graphicx,color}
\usepackage{graphicx}
\usepackage{dcolumn}
\usepackage{bm}

\begin{document}
\preprint{cond-mat}
\title{Suppression of electronic susceptibility in metal-Mott insulator alternating material,  (Me-3,5-DIP)[Ni(dmit)$_2$]$_2$}
\author{S.\ Fujiyama}
\email{fujiyama@ap.t.u-tokyo.ac.jp} 
\affiliation{Department of Applied Physics, University of Tokyo, Tokyo 113-8656, Japan}
\affiliation{CREST, Japan Science and Technology Corporation, Kawaguchi 332-0012, Japan}
\author{A.\ Shitade}
\affiliation{Department of Applied Physics, University of Tokyo, Tokyo 113-8656, Japan}
\author{K.\ Kanoda}
\affiliation{Department of Applied Physics, University of Tokyo, Tokyo 113-8656, Japan}
\affiliation{CREST, Japan Science and Technology Corporation, Kawaguchi 332-0012, Japan}
\author{Y.\ Kosaka}
\author{H.\ M.\ Yamamoto}
\author{R.\ Kato}
\affiliation{CREST, Japan Science and Technology Corporation, Kawaguchi 332-0012, Japan}
\affiliation{RIKEN, Wako-shi, 351-0198, Japan}
\date{\today}
\begin{abstract}
Frequency shifts and nuclear relaxations of $^{13}$C NMR of the metal-insulator alternating material, (Me-3,5-DIP)[Ni(dmit)$_2$]$_2$, are presented. The NMR absorption lines originating from metallic and insulating layers are well resolved, which evidences the coexistence of localized spins ($\pi_\textrm{loc}$) and conduction $\pi$-electrons. The insulating layer is newly found to undergo antiferromagnetic long range order at about 2.5 K, suggesting emergence of $S=1/2$ Mott insulator. In the metallic layer, we found significant suppressions of static and dynamical susceptibilities of conduction electrons below 35 K, where antiferromagnetic correlation in the insulating layer evolves. We propose a dynamical effect through strong $\pi$-$\pi_\textrm{loc}$ coupling between the metallic and insulating layers as an origin of the reduction of the density of states.
\end{abstract}

\maketitle
Correlations between conduction electrons and localized spins attract much attention in condensed matter physics. Competition between magnetic ordering due to the Ruderman-Kittel-Kasuya-Yosida (RKKY) interaction and the formation of a Kondo singlet accompanied by mass enhancement of conduction electrons is a constitutive problem~\cite{Tsunetsugu1997}. In the last decades, several organic charge-transfer-salts with anions containing magnetic $d$ ions have been synthesized~\cite{Day1992,Coronado2000}. It is widely recognized that physical properties of molecular conductors 
are determined by a single band formed by the frontier orbital consisting of hybridized $\pi$-electons on molecules, and simple tight binding approximation works extremely 
well for the description of the band. Such simplicity is a 
significant advantage in the study of $\pi$-$d$ correlation between conduction 
$\pi$-electrons and local $d$ moments.

Since hitherto known organic charge-transfer-salts with $\pi$-$d$ interaction contain transition metals as localized spins, the wave functions of conduction electrons and localized spins in the anions are weakly hybridized because of large energy gap between the HOMO and the $d$ levels up to 1 eV~\cite{Mori2002,Zaanen1987}. In this case, $\pi$-$d$ interaction is well described by mean field approximation where conduction electrons feel exchange fields from local $d$ moments~\cite{Hotta2000,Mori2002,Fujiyama2006b}.

\begin{figure}[htb]
\includegraphics*[width=6cm]{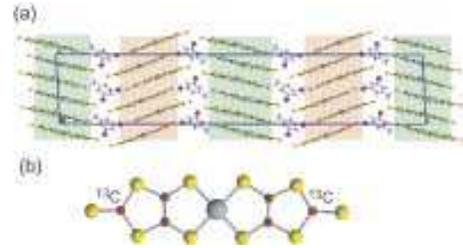}
\caption{(a) Crystal structure of (Me-3,5-DIP)[Ni(dmit)$_2$]$_2$. The metallic and insulating Ni(dmit)$_{2}$ layers are hatched by different colors. (b) Structure of Ni(dmit)$_{2}$. Two carbon atoms at both ends of the molecule are labeled by $^{13}$C (colored by red).}
\label{fig:structure}
\end{figure}

A newly synthesized organic conductor, (Me-3,5-DIP)[Ni(dmit)$_2$]$_2$, is a candidate material to study the correlation between conduction $\pi$-electrons and localized $\pi$-spins. This material consists of two dimensional quarter-filled Ni(dmit)$_2$ anion layers and nested (Me-3,5-DIP) cations as shown in Fig.\ \ref{fig:structure}. There are two inequivalently aligned Ni(dmit)$_{2}$ layers alternating in a unit cell. An extended H\"uckel calculation expects an oval Fermi surface originating from one Ni(dmit)$_{2}$ layer. In the other Ni(dmit)$_{2}$ layer, strong dimerization is expected originating from 30 times as large intradimer transfer integral as interdimer one, which leads to a half-filled band~\cite{Kosaka2007}.

\begin{figure}[htb]
\includegraphics*[width=6cm]{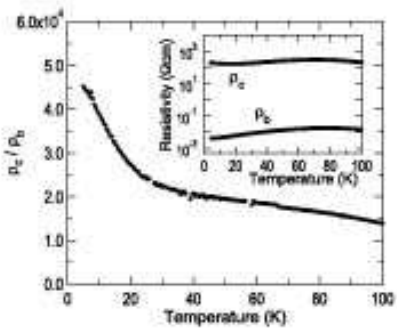}
\caption{Anisotropy of resistivities. The raw data of $\rho_{\textrm{b}}$ and $\rho_{\textrm{c}}$ are plotted in the inset.}
\label{fig:ResAniso}
\end{figure}

Experimentally, the results of macroscopic measurements support alternation of metallic and Mott insulating layers~\cite{Kosaka2007}. We plot in Fig.\ \ref{fig:ResAniso} the anisotropy of resistivity. The ratio of inter- and intra- layer resistivities, $\rho_c/\rho_b$, which is 50 at room temperature, increases upon cooling and exceeds 1000 below 180 K, indicating a growth of strong two-dimensional anisotropy. As for the magnetism, the susceptibility above 40 K can be fitted to a summation of two components, $\chi(T)=1/2(\chi_{\mathrm{Curie}}(T)+\chi_{\mathrm{Pauli}})$, where $\chi_{\mathrm{Curie}}(T)=C/(T-\Theta)$($C=0.375$ emu K mol$^{-1}$, $\Theta=-5$ K) and $\chi_{\mathrm{Pauli}}=7.2 \times 10^{-4}$emu mol$^{-1}$~\cite{Kosaka2007}. This suggests a coexistence of $S=1/2$ localized spins with antiferromagnetic correlations and Pauli paramagnetic conduction electrons.

In contrast to known organic conductors with $\pi$-$d$ interaction, $S=1/2$ localized spins ($\pi_{\textrm{loc}}$) are discussed to emerge within Ni(dmit)$_{2}$ molecular layers in the present material. This prominent feature leads to (i) widely distributed spin density because of $\pi$ electron character and small energy gap between conduction electrons and localized $\pi_{\textrm{loc }}$ spins, (ii) large interlayer transfer integrals ($\sim 3$ meV) comparable to inplane interdimer transfers ($< 3.4$ meV) as is estimated by extended H\"uckel calculation~\cite{Kosaka2007}, (iii) possible quantum effect owing to $S=1/2$ localized spins. Moreover, we can coordinate indivisual microscopic measurements of $\pi$-$\pi_\textrm{loc}$ correlation by labeling carbons both in the metallic and insulating Ni(dmit)$_{2}$ layers by NMR-active $^{13}$C. This has a distinct advantage to investigate correlations between conduction electrons and localized spins, which has never been exploited in either rare earth compounds or $\pi$-$d$ organic conductors.

In this Letter, we report frequency shifts and nuclear relaxations of $^{13}$C NMR of the metal-insulator alternating material, (Me-3,5-DIP)[Ni(dmit)$_2$]$_2$. The NMR absorption lines originating from both layers are well resolved, which evidences the coexistence of metallic and insulating components. The NMR line of the insulating layers shows a significant broadening originating from antiferromagnetic correlation below about 35 K. And finally, the insulating layers are found to undergo antiferromagnetic long range order at about 2.5 K. This supports the argument based on the extended H\"uckel calculation that the insulating Ni(dmit)$_{2}$ layers are in a Mott insulating state. In the metallic layer, significant decreases of the frequency shift, $1/T_{1}T$, and $1/T_{2G}$ are observed below 35 K, which show suppressions of the static and dynamical susceptibilities indicating an anomalous reduction of the density of states (DOS) at the Fermi level. We propose a dynamical effect of short range ordering of localized $\pi_\textrm{loc}$ spins on conduction electrons through strong $\pi$-$\pi_\textrm{loc}$ hybridization as an origin of this anomalous reduction of the DOS.

We used a plate-like single crystal of (Me-3,5-DIP)[Ni(dmit)$_2$]$_2$ synthesized by electrochemical reaction. Carbon atoms at both ends of a Ni(dmit)$_{2}$ molecule are labeled by $^{13}$C. External fields were applied in the $c$ direction (perpendicular to the Ni(dmit)$_{2}$ layers). In this condition, two $^{13}$C atoms in a Ni(dmit)$_{2}$ molecule are inequivalent because of dimerization of the molecules in both layers.

\begin{figure}[htb]
\centering
\includegraphics*[width=6.5cm]{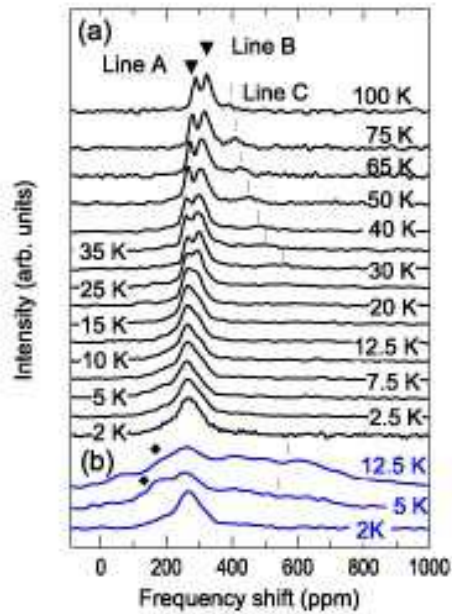}
\caption{(a) $^{13}$C NMR spectra under the condition of nuclear magnetization fully polarized. Distinct lines are labeled as A, B, and C. (b) Spectra obtained by reducing the time interval between saturating and $\pi/2$ pulses to 1/15 of $T_{1}$ for Lines A and B. At 2 K, Line C is missing and only the residual component of Lines A and B is observed.}
\label{fig:Spectra}
\end{figure}

The $^{13}$C NMR spectra above 35 K are composed of two distinct lines (Lines A and B) and one broad and small line (Line C) as shown in Fig.\ \ref{fig:Spectra}. While the peak frequencies and linewidth of Lines A and B are nearly independent of temperature ($T$), those of Line C strongly depend on $T$. Moreover, the nuclear spin-lattice relaxation rate, $1/T_{1}$, of Line C is about 20 times as large as those of Lines A and B. 
Since the Ni(dmit)$_{2}$ molecules stack differently in the insulating and metallic layers, we expect four $^{13}$C lines in the spectra. We found one obscured line of which the linewidth and $1/T_{1}$ are similar to those of Line C at about 150 ppm as marked by diamonds in Fig.\ \ref{fig:Spectra} (b), by saturating nuclear magnetization so as to suppress signal intensities of Lines A and B.

\begin{figure}[htb]
\centering
\includegraphics*[width=6.5cm]{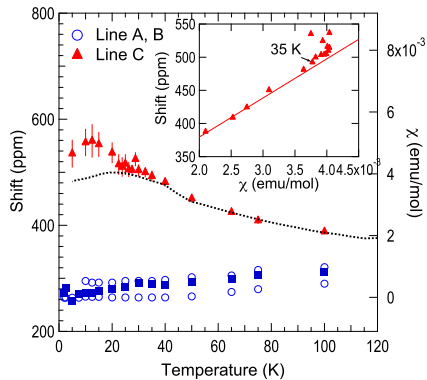}
\caption{Frequency shifts of Lines A, B, and C. The peak positions ({\color{blue}$\circ$}) below 20 K are defined by deconvolution of the spectra. Center of gravities of Lines A and B ($K_{\mathrm{A}}$, {\color{blue}$\blacksquare$}) and Line C ($K_{\mathrm{C}}$, {\color{red}$\blacktriangle$}) are plotted. The dotted line is the bulk susceptibility measured by SQUID scaled by $A_{\parallel}\chi(T)/N_{A}\mu_{B}$.}
\label{fig:Shift}
\end{figure}

We plot in Fig.\ \ref{fig:Shift} the frequency shifts, $K$, (peak positions and the center of gravities) of Lines A, B, and C.
The $K_\mathrm{C}$ shows Curie-Weiss-like $T$ dependence, and follows the susceptibility measured by SQUID above 35 K as shown in the inset of Fig.\ \ref{fig:Shift}. Therefore, Line C is assigned to the insulating layers, while Lines A and B are to the metallic layers. The hyperfine coupling constant is estimated as  $A_{\parallel}(\mathbf{q}=0)=K_{\mathrm{C}}(T) N_{A}\mu_{B}/\chi(T)=410$ (Oe/$\mu_B$). This value of $A_{\parallel}$ is less than 0.1 times as that of the central carbons in BEDT-TTF or TMTTF donors, which shows much smaller spin density at the end carbons in Ni(dmit)$_{2}$ molecules~\cite{DeSoto1995}.

It should be noted that $K_{\mathrm{C}}$ exceeds the straight line of the slope of $A_{\parallel}$ below 35 K as shown in the inset of Fig.\ \ref{fig:Shift}. If the susceptibility from the metallic layers were independent of $T$, the linear relation between $K_{\mathrm{C}}$ and the susceptibility would hold in the paramagnetic state. Therefore, this deviation of $K_{\mathrm{C}}$ suggests a reduction of the local susceptibility below 35 K in the metallic layer.
Indeed, the slope of $K_{\mathrm{A}}$ changes below 35 K, indicating the suppression of conduction electron susceptibility.

\begin{figure}[htb]
\centering
\includegraphics*[width=5.5cm]{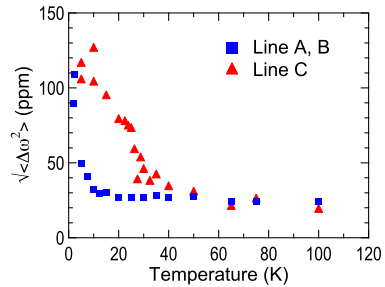}
\caption{Second moments of Lines A, B, and C.}
\label{fig:SecMom}
\end{figure}

We plot second moments of distinct lines of the spectra, $\sqrt{\langle \Delta \omega ^{2} \rangle}$, in Fig.\ \ref{fig:SecMom}. Line C shows a significant broadening below 35 K, indicating a development of antiferromagnetic correlation. The linewidth grows down to 5 K; however, Line C disappears below 2.5 K, corresponding to antiferromagnetic long range order in the insulating layer as shown in Fig.\ \ref{fig:Spectra}(b). The observed transition temperature $T_\mathrm{N}<5$ K is much lower than 35 K where the antiferromagnetic correlation sets in, possibly due to quasi-one-dimensional anisotropy of magnetic interaction $J$ estimated by the extended H\"uckel calculation~\cite{Kosaka2007}.
Moderate evolution of $\sqrt{\langle \Delta \omega ^{2} \rangle}$ of Lines A and B is also observed below about 50 K, and the line broadening becomes appreciable below 15 K. Below 2.5 K, both lines are strongly broadened because of antiferromagnetic order in the insulating layer. The estimated amplitude of the ordered moments are 0.2 $\mu_{B}$ at 2 K by assuming dipolar coupling between nuclear carbons in the metallic layer and the localized spins.
It should be stressed that this microscopic evidence for antiferromagnetic ordering in the insulating layer strongly supports the emergence of $S=1/2$ localized spins at dimerized Ni(dmit)$_{2}$ molecules and the argument that a Mott insulating state is realized in the insulating layers.

\begin{figure}[htb]
\centering
\includegraphics*[width=6cm]{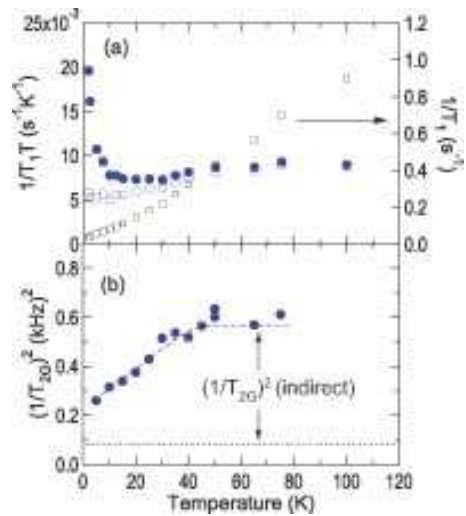}
\caption{(a) $1/T_{1}T$ ($\bullet, \circ$), $1/T_{1}$ ($\Box$, right axis) and (b) $1/T_{2G}$ of Lines A and B. The dashed line shows the level of $\langle \Delta\omega^{2}_{\mathrm{direct}}\rangle$ estimated by Eq.~\ref{eq:T2}.}
\label{fig:T1}
\end{figure}

We plot nuclear spin-lattice ($1/T_{1}T$) and spin-spin ($1/T_{2G}$) relaxations of the metallic layers in Fig.\ \ref{fig:T1}.
The $1/T_{1}T$ is enhanced below 15 K, which is consistent with the increase of $\sqrt{\langle \Delta\omega \rangle ^{2}}$ as shown in Fig.\ \ref{fig:SecMom}. This observation suggests an enhancement of the magnetic correlation at $\mathbf{q}\neq 0$ of the conduction electrons induced by antiferromagntic fluctuation of the localized spins in the insulating layers. However, no evidence of magnetic long range order of conduction electrons is suggested even below $T_\mathrm{N}\sim$2.5 K because no critical slowing down is found in $1/T_{1}$. When we fit $1/T_{1}T$ below 15 K to a Curie-Weiss formula, $\Theta=-5.7$ K is obtained. 
We subtract the $y$-intercept of $1/T_{1}$ at $T=0$ from the raw data of $1/T_{1}$ and plot $1/T_{1}T$ with open circles as a trial to exclude the low temperature enhancement originating from the critical slowing down of the localized spins in the insulating layers. As shown in Fig.\ \ref{fig:T1} (a), $1/T_{1}T$ is nearly independent of $T$ above 40 K, as is expected in metals with weak electronic correlations. Below 35 K, however, an evident reduction of $1/T_{1}T$ is observed upon cooling. The $1/T_{1}T$ of metals is related to the DOS at the Fermi level by $1/T_1T=2\gamma_{n}^{2}/\mu_{B}^{2}\sum_{q}A_{\perp}(\mathbf{q})^{2}\chi''(\mathbf{q}, \omega)/\omega \propto D(E_F)^{2}$; therefore, our observation of the reduction of $1/T_{1}T$ suggests an about 20 \% reduction of the DOS.

This argument is supported by the behavior of nuclear spin-spin relaxation rate, $1/T_{2G}$, which is proportional to the real part of the dynamical susceptibility of conduction electrons. The spin echo decay curves are well fitted by Gaussian-Lorenzian formula, $M(2\tau)/M_{0}=\exp[-(2\tau/T_{2L}+2(\tau/T_{2G})^{2})]$, for the whole temperature range. Here, $T_{2L}$ ($T_{2G}$) is the Lorenzian (Gaussian) component of the echo decay time and $\tau$ is the time interval between $\pi/2$ and $\pi$ pulses. No significant enhancement of $1/T_{2L}$ even below $T_\mathrm{N}$ again supports the absence of the critical slowing down of conduction electron spins. Concerning the Gaussian decay, the RKKY-type indirect spin-spin coupling mediated by conduction electrons as well as direct nuclear dipolar interactions contribute to $1/T_{2G}$ as, 
\begin{eqnarray}
\left(\frac{1}{T_{2G}}\right)^{2}=\langle \Delta \omega^{2}_\mathrm{direct} \rangle+\sum_{q}A(\mathbf{q})^{2}\chi'(\mathbf{q}, \omega=0)^{2}, \\
\langle \Delta \omega^{2}_\mathrm{direct}\rangle=\frac{3}{4}\gamma_{n}^{4}\hbar^{2}I(I+1)\sum_{k}\frac{(1-3\cos^{2}\theta_{jk})^{2}}{r_{jk}^{6}}.
\label{eq:T2}
\end{eqnarray}
Here, $\langle \Delta \omega^{2}_\mathrm{direct} \rangle$ is independent of $T$ and calculated as $0.09$(kHz$^{2}$), which is plotted by a dashed line in Fig.~\ref{fig:T1}~(b). Since $(1/T_{2G})^{2}$ subtracted by $\langle \Delta \omega^{2}_\mathrm{direct} \rangle$ is proportional to $\chi'(\mathbf{q},\omega=0)^{2}$, the reduction of $1/T_{2G}$ below 35 K suggests about a 30 \% reduction of the DOS.

The reductions of static and dynamical susceptibilities indicating the suppression of the DOS below 35 K are unconventional for metals with weak electronic correlations. A similar suppression of the susceptibility has been reported in other two-dimensional metals such as underdoped high-$T_{\mathrm{c}}$ cuprates and $\kappa$-(BEDT-TTF)$_{2}$\textit{X} ($\kappa$-ET) organic conductors, which is known as the pseudogap phenomenon~\cite{Yasuoka1989,Kanoda1997}. There, the reduction of the DOS also causes the increase of two dimensional anisotropy in the resistivity below the temperature where the pseudogap appears~\cite{Ito1993}. In the present compound, as shown in Fig.~\ref{fig:ResAniso}, we also observe an anomaly in the anisotropy of resistivity at about 35 K, below which the anisotropy increases rapidly, in accordance with that in underdoped high-$T_{\mathrm{c}}$ cuprates. In spite of these common features, the origin of the suppression of the DOS in the present material appears to be different from that in cuprates or $\kappa$-ET. In cuprates or $\kappa$-ET, the antiferromagnetic correlation develops in conduction electrons. On the contrary, in the present material, the antiferromagnetic correlation grows dominantly in the insulating layers, not in the metallic layers, as is evidenced by the broadening of Line C in Fig.~\ref{fig:SecMom}: DOS is suppressed through a strong coupling between the $S$=1/2 moments in the insulating layers and the conduction electrons in the metallic layers. In fact, the extended H\"uckel calculation supports the existence of this strong $\pi$-$\pi_\textrm{loc}$ coupling: the estimated interlayer transfer integral is comparable to the interdimer one within the layer~\cite{Kosaka2007}. Furthermore, our estimate of the Weiss constant, $\Theta=-5.7$ K, based on $1/T_{1}$ for conduction electrons manifests the magnetic correlation induced by the short range order in the adjacent insulating layers through this $\pi$-$\pi_\textrm{loc}$ interaction. A coupling between localized spins and conduction electrons has also been observed in the hitherto known $\pi$-$d$ organic materials~\cite{Fujiyama2006b}. The effect of the $\pi$-$d$ interaction is, however, well described by mean field analysis, and does not lead to the dynamical effect observed here. Hence, the suppression of the DOS that we found here is characteristic to the strong $\pi$-$\pi_\textrm{loc}$ interaction in the present compound, and its origin is distinct from ones in the cuprates, $\kappa$-ET and $\pi$-$d$ organic conductors.

Let us discuss how the antiferromagnetic correlation in the insulating layers induces the suppression of the DOS in the metallic layers. A possible source is the instability toward the topological change of the Fermi surface through the antiferromagnetic correlation of the localized spins. When the insulating layers are antiferromagnetically ordered, DOS can be reduced by folding of the Fermi surface through doubling of Brillouin zone. A strong hybridization between $\pi$ and $\pi_\textrm{loc}$ spin might give rise to the fluctuation of the folding of the Fermi surface even above $T_{\mathrm{N}}$, leading to the suppression of the DOS.

As a summary, our observation of two constituent lines of which the frequency shifts and linewidth differently depend on $T$ in $^{13}$C NMR spectra evidences that (Me-3,5-DIP)[Ni(dmit)$_2$]$_2$ is a hybrid material of metallic and insulating Ni(dmit)$_{2}$ layers. We found an antiferromagnetic order in the insulating layers at $T_\mathrm{N}\sim 2.5$ K, suggesting $S=1/2$ Mott insulator through dimerization of Ni(dmit)$_{2}$ molecules.
In the metallic layers, we found about 20 \% reductions of static and dynamical susceptibilities of conduction electrons below 35 K, suggesting the reduction of the DOS at the Fermi level. This characteristic behavior is seen below the temperature at which the antiferromagnetic correlation starts to grow in the insulating layers. We propose a dynamical effect through strong $\pi$-$\pi_\textrm{loc}$ hybridization as an origin of the observed anomalous reduction of the DOS. We now stand at a point where electronic correlation between conduction electrons and localized $S=1/2$ spins are investigated in respective channels of metallic and insulating layers separately, which has been impossible in rare earth intermetallic compounds as well as $\pi$-$d$ organic conductors.

\begin{acknowledgments}
We are grateful to Y.\ Motome, Y.\ Itoh, J.\ Kishine and H.\ Seo for fruitful discussions. This work is supported by the JSPS (No.\ 17740213 and 19740200) and the MEXT (No.\ 13640375, 16076204, and 16GS0219).
\end{acknowledgments}

\end{document}